\documentclass{ws-jai}
\usepackage[flushleft]{threeparttable}
\usepackage{subcaption}
\usepackage{adjustbox}
\usepackage[dvipsnames]{xcolor}

\begin{document}

\catchline{}{}{}{}{} 

\markboth{Doaa Eid., et al}{A Hybrid Genetic-Fuzzy Controller for a 14-inches Astronomical Telescope Tracking}

\title{A Hybrid Genetic-Fuzzy Controller for a 14-inches Astronomical Telescope Tracking\\}

\author{Doaa Eid
$^{1,*}$, Abdel-Fattah Attia$^{2}$, Said Elmasry$^{3}$ and Islam Helmy$^{1}$}

\address{
$^{1}$Astronomy Department National Research Institute of Astronomy and Geophysics, Helwan, 11421, Egypt\\
$^{2}$Electrical Engineering Department, Faculty of Engineering, Kafrelsheikh University, Kafrelsheikh, 33511, Egypt\\
$^{3}$Department of Electrical Power Engineering, Faculty of Engineering, Helwan University, Cairo, 11792, Egypt}

\maketitle

\corres{$^{*}$Corresponding author: Doaa Eid \\ e-mail: doaa\_eid@nriag.sci.eg.}

\begin{history}
\received{(to be inserted by publisher)};
\revised{(to be inserted by publisher)};
\accepted{(to be inserted by publisher)};
\end{history}

\begin{abstract}
The performance of on telescope depend strongly on its operating conditions. During pointing the telescope can move at a relatively high velocity, and the system can tolerate trajectory position errors higher than during tracking. On the contrary, during tracking Alt-Az telescopes generally move slower but still in a large dynamic range. In this case, the position errors must be as close to zero as possible. Tracking is one of the essential factors that affect the quality of astronomical observations. In this paper, a hybrid Genetic-Fuzzy approach to control the movement of a two-link direct-drive Celestron telescope is introduced. The proposed controller uses the Genetic algorithm (GA) for optimizing a fuzzy logic controller (FLC) to improve the tracking of the 14-inches Celestron telescope of the Kottamia Astronomical Observatory (KAO). The fuzzy logic input is a vector of the position error and its rate of change, and the output is a torque. The GA objective function used here is the Integral Time Absolute Error (ITAE). The proposed method is compared with a conventional Proportional-Differential (PD) controller, an optimized PD controller with a GA, and a Fuzzy controller. The results show the effectiveness of the proposed controller to improve the dynamic response of the overall system.
\end{abstract}


\section{Introduction}
\noindent The Telescope is a far distance direct viewing system. It is used to map an angular pattern of the sky onto a detector by collecting photons. The astronomical telescopes are a precise and expensive piece of equipment. The accuracy of pointing and tracking of astronomical telescope critically affect the quality of the astronomical observations. Consequently, the pointing and tracking processes are needing sophisticated means of control. 

The Celestron telescope dynamic equations are a set of nonlinear-coupled differential equations that involve high nonlinearities, strong coupling, and uncertain system dynamics.  Various adaptive control strategies have been suggested to overcome the linearization control drawbacks. Many of these strategies generally depend on neglecting friction, backlash, and other unmodeled dynamics \cite{A.Attia97, F. Lewis93}. 

It is known that conventional controllers like PD controllers generally could not work well for nonlinear systems, higher-order and time-delayed linear systems, and particularly systems that have no precise mathematical models.  An auto-tuning P D controller was developed to get rid of these defects \cite{S. Deif11}. Moreover, it is recommended to make use of an Artificial Intelligent based controller.

Fuzzy logic control \cite{L. Zadeh65} has given great potential. Since it has been able to handle uncertainty, nonlinearity using the programming capability of human control behavior. The fuzzy system has some challenges, for instance, the determinations of shapes, parameters of membership functions, and fuzzy-rules. It is conventionally tuned using trial and error by operator. Thus, time and effort are consumed, and the optimal fuzzy system is more difficult to be designed. Several optimization methods have been presented to solve these challenges, such as the Genetic algorithm \cite{Koji Shimojima95}.

Genetic algorithms (GAs) \cite{D. Goldberg89} are computational strategies that rely on the processes of natural evolution. The used operators realize a process of heuristic-search in a search space.

The main focus of this paper is to study the performance of the proposed controller. The suggested approach is to apply the Genetic algorithm to optimize the Fuzzy controller for the Celestron telescope to get an output with better dynamics and static performance. Besides, an investigation of an optimized PD controller with the Genetic algorithm. The best controller will be used to develop an automatic tracking for the 14-inches Celestron telescope of the Kottamia Astronomical Observatory (KAO).

This paper is organized as follows: Section \ref{sec.2} provides the model of the 14-inches Celestron telescope. Section \ref{sec.3} introduces a description of the compared controllers. Section \ref{sec.4} describes the proposed controller. Finally, Sections \ref{sec.5} and \ref{sec.6} illustrate the experimental results and conclusion, respectively.

\section{Celestron Telescope Dynamics and Modeling}\label{sec.2}
\begin{figure}[h]
\begin{center}
\includegraphics{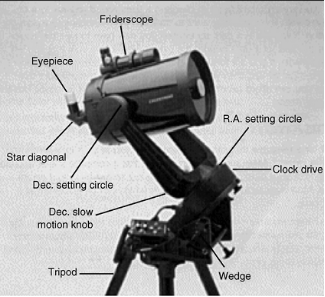} 
\end{center}
\caption{The 14-inches Celestron telescope.}
\label{fig1}
\end{figure}

The Celestron telescope is moved in two directions, as presented in Figure \ref{fig1}. The Right Ascension (RA) coordinates and the Declination (DEC) have two motor drives of the telescope movements on both sides \cite{Celestron92}.

The dynamical analysis of the telescope investigates a relation between the joint torques/inertias applied by the actuators and the position, velocity, and acceleration of the telescope tube with respect to the time \cite{Ouamri Bachira12}.

The dynamic equations of the telescope are represented by the following coupled non-linear differential equations:

\begin{equation}
 M (\theta) \, \ddot {\theta} + N (\theta , \dot{\theta}) + \tau_d = \tau
.\label{eq1}
\end{equation}

\begin{equation}
N(\theta , \dot{\theta})= C (\theta , \dot{\theta}) + G (\theta )
.\label{eq3}
\end{equation}

$ \tau_d$ is a constant disturbance torque  \cite{A.Attia04}, which represents the unknown dynamics, such as friction and other disturbances. In the case of absence, we can neglect it \cite{V. Santibaiiez98}. By substituting equation \ref{eq3} in equation \ref{eq1}, the equation will be as shown in equation \ref{eq2}.
\begin{equation}
M (\theta) \, \ddot {\theta}+ C (\theta , \dot{\theta}) + G (\theta ) = \tau
.\label{eq2}
\end{equation}

 Where :
$$\theta=[\theta_1 \, \theta_2]^T, \dot{\theta}=[\dot{\theta}_1  \, \dot{\theta}_2]^T,  \ddot {\theta} = [\ddot {\theta}_1 \, \ddot {\theta}_2]^T$$

Where $M (\theta )$   is the 2 x 2 inertia matrix,  $ C (\theta , \dot{\theta}) $ is the 2 x 1 Coriolis/Centrifugal vector, and $G (\theta )$  is the 2 x 1 Gravitational torque vector, and $\tau$ is Torque. Besides, $\theta$, $\dot{\theta}$, and $ \ddot {\theta} $ are the joint angular position, velocity, and acceleration terms, respectively.  Also, $\tau_d$ is the vector of dynamic and static friction forces. The gravity terms $G$ tends to zero because the telescope must be well-balanced using some counterweights \cite{A.Attia97}.

The telescope model is decoupled using a compensator. The decoupled system consists of independent variable systems. Thus, the compensated telescope model represents a linear system, which allows using the linear PD controller for control \cite{A.Attia04}.

\section{Investigated Controllers}\label{sec.3}
\subsection{Conventional PD Controller}\label{sec.3.1}

The PD computed-torque controller with an auxiliary control signal is selected to control the telescope.
This controller has duplicated for each arm of the telescope. The control action at each arm is designed to be independent of the control action at the other arm, as shown in Figure \ref{fig2}. This type of control is called independent-joint control. The output (torque) of the PD controller is described using the following equation:

\begin{figure}[h]
\begin{center}
\includegraphics[width=16cm,height=8cm]{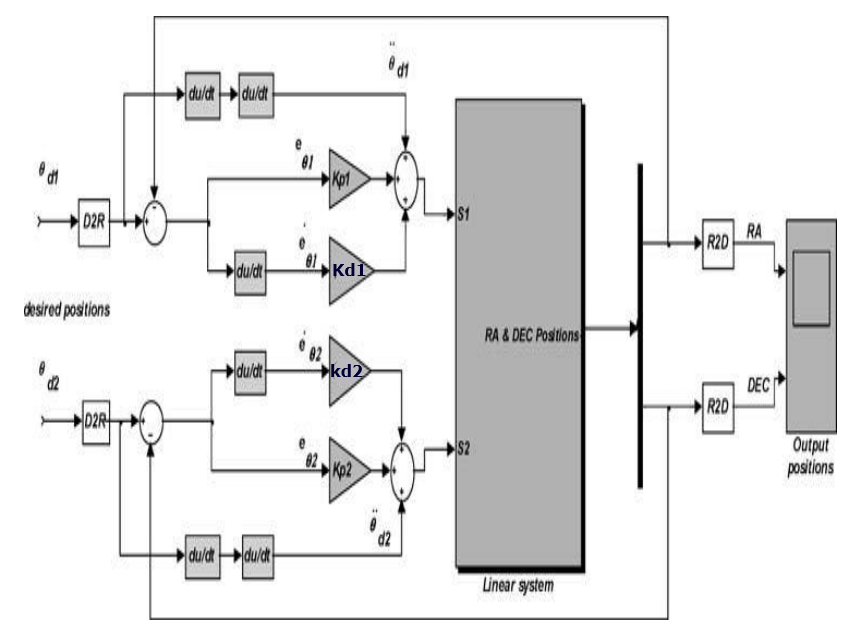} 
\end{center}
\caption{PD computed-torque controller for the telescope model.}
\label{fig2}
\end{figure}

\begin{equation}
\tau= M (\theta)(\ddot {\theta}_d + K_d \dot{e}+K_pe)+C (\theta , \dot{\theta})
\label{eq4}
\end{equation}

Where $k_p$, $k_d$ are the proportional and derivative gains of the PD controllers, respectively. Also,  $e (\theta )$ is the difference between the desired angle position  and the actual one of the Celestron in a certain direction. The optimal gains of the PD controllers are determined by making use of the Ziegler-Nichols rule.

\subsection{Fuzzy Logic Controller (FLC) Design}\label{sec.AssReslts}
\begin{figure}[h]
\centering
\begin{subfigure}{1.0\textwidth}
  \centering
  \includegraphics[width=12cm,height=5cm]{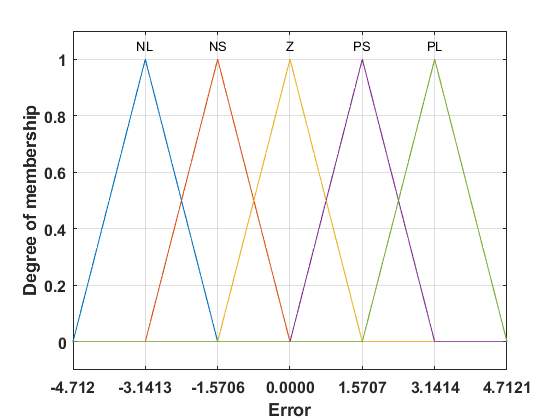}
  \caption{Memberships of the error.}
  \label{fig3a}
\end{subfigure}%
\vspace*{8pt}
\begin{subfigure}{1.0\textwidth}
  \centering
  \includegraphics[width=12cm,height=5cm]{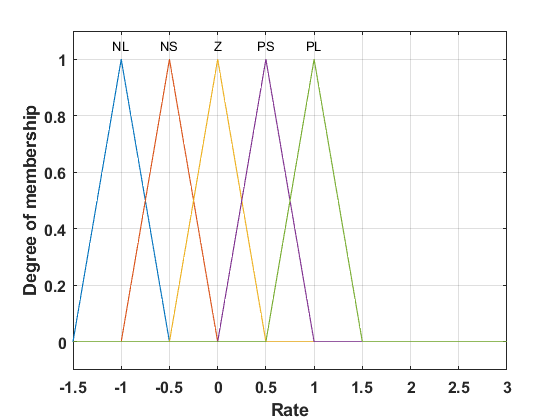}
  \caption{Memberships of the error rate of change.}
  \label{fig3b}
\end{subfigure}

\begin{subfigure}{1.0\textwidth}
  \centering
  \includegraphics[width=12cm,height=5cm]{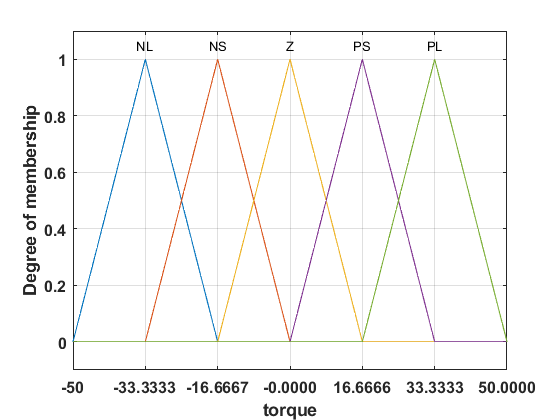}
  \caption{Memberships of the torque.}
  \label{fig3c}
\end{subfigure}%
\caption{Membership functions of the fuzzy inputs and output. (a) : Memberships of the error. (b) : Memberships of the error rate of change. (c) : Memberships of the torque. }\label{fig3}
\end{figure}

Two inputs and a single output fuzzy controller has used \cite{A.Attia04}. The inputs are position and velocity error, and the output is a torque used to move the telescope-arm. The position error is the difference between the desired position and the telescope one, while the velocity error is the rate of change of the position error with the time. The inputs and the output membership functions are triangular, as presented in Figure \ref{fig3a}, \ref{fig3b}, and \ref{fig3c}, respectively.

With the above membership functions, the inference composition rules are defined as follows in Table \ref{tbl1}. The fuzzy logic output (Defuzzification) relies on the center of area (COA) method.\par

\begin{wstable}[h]
\caption{Look-up table of $e_{\theta1}$, and $\dot{e}_{\theta1}$.}
\begin{tabular}{@{}cccccc@{}} \toprule
$e_{\theta1}$/$\dot{e}_{\theta1}$ & NL & NS & Z & PS & PL \\ 
\colrule
NL\hphantom{00} & NL\hphantom{0} & NL\hphantom{0} & NL\hphantom{0} & NS\hphantom{0} & Z\hphantom{0} \\
NS\hphantom{00} & NL\hphantom{0} & NL\hphantom{0} & NS\hphantom{0} & Z\hphantom{0} & PS\hphantom{0} \\
Z\hphantom{0} & NL\hphantom{0} & NS\hphantom{0} & Z\hphantom{0} & PS\hphantom{0} & PL\hphantom{0} \\
PS\hphantom{0} & NS\hphantom{0} & Z\hphantom{0} &PS\hphantom & PL\hphantom{0} & PL\hphantom{0} \\
PL\hphantom{0} & Z\hphantom{0} & PS\hphantom{0} & PL\hphantom{0} & PL\hphantom{0} & PL\hphantom{0} \\
\botrule
\end{tabular}
\label{tbl1}
\end{wstable}

\section{Proposed Controllers}\label{sec.4}
\subsection{GA-PD  Controller}\label{GA-PD}
\begin{figure}[h]
\begin{center}
\includegraphics[width=16cm,height=8cm]{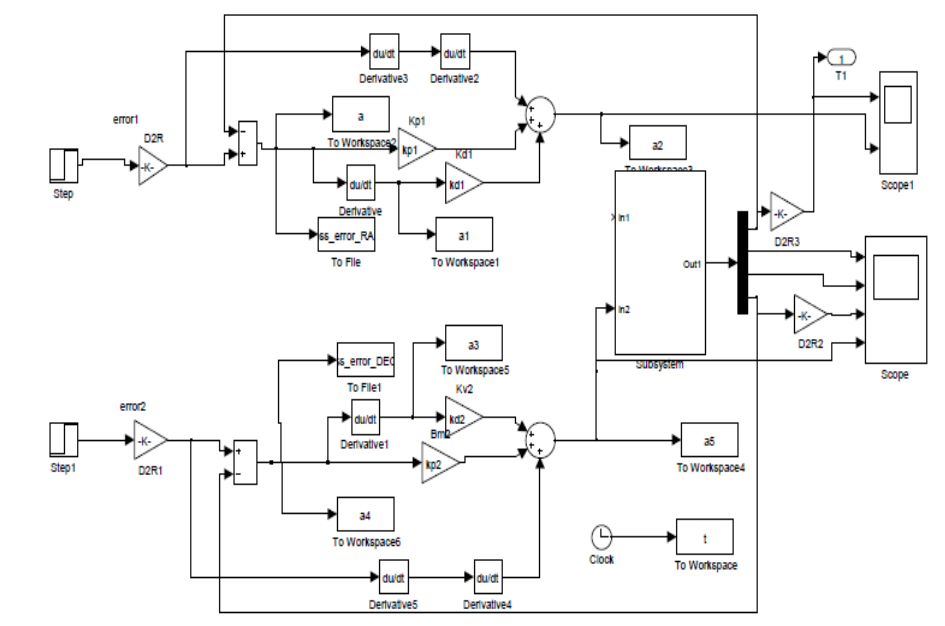} 
\end{center}
\caption{GA-PD controller for the telescope model.}
\label{fig5}
\end{figure}

The Genetic algorithm (GA) is a powerful searching method. Besides, it is a numerical optimization algorithm based on natural selection. GA is an effective method to solve optimization problems, and it is superior in avoiding local minima, which is a common side in a nonlinear system. GA starts with an initial population containing some of the chromosomes. Each chromosome represents a problem solution. Its performance is evaluated by a fitness function.
PD controller has become preferable in control of industries because of its simplicity and effectiveness, but a real demand lies in tuning them to meet the expectations. GA has the ability to find optimal solutions in complex search space. It has been applied to PD controllers to design the controller gains. Figure \ref{fig5} presents a detailed block diagram for Genetic tuning PD controller for telescope model.

The fitness function plays a principal role in the genetic algorithm because the search direction depends on it.  Minimize output error is one of the essential aims of control systems. Integral Time Absolute Error (ITAE) is used as a fitness function, as seen in equation \ref{eq6}.
Where $e(t)$  is the position error, and $\alpha$, $\beta$ are the weighting coefficients, such that $\alpha=\beta=0.01$.

\begin{equation}\label{eq6}
ITAE = \int_{0}^{T}(\alpha\left| e(t) \right|+\beta\left| \acute{e(t)} \right|)dt
\end{equation}

\subsection{GA-FLC Controller}

\begin{figure}[h]
\begin{center}
\includegraphics[width=16cm,height=7cm]{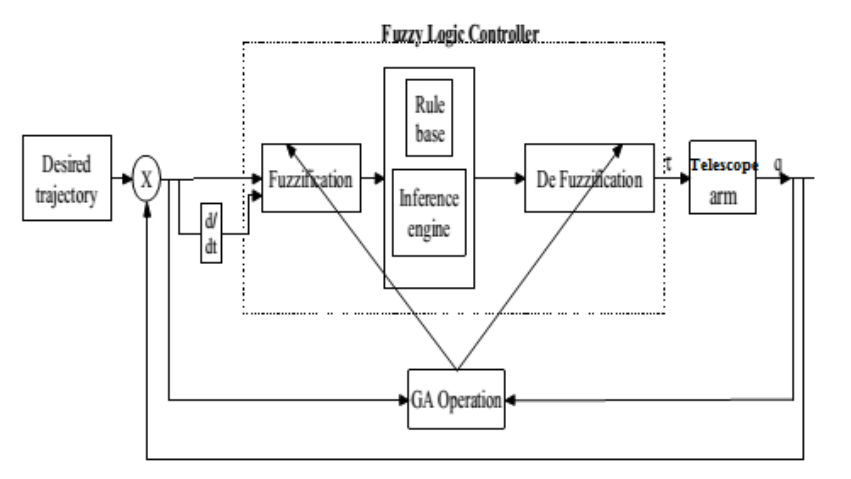} 
\end{center}
\caption{Structure of Genetic tuned fuzzy control system.}
\label{fig6}
\end{figure}

Fuzzy controller design relies on expert knowledge of a system, which is intended to control it. So control engineers must have a great experience with the system. The fuzzy controller is often designed by try and error, which is not simple, as well as it is time-consuming.  That is because several parameters affect the fuzzy controller, such as membership shapes, inference rules, inputs gain, and outputs gain. In this paper, the fuzzy membership functions are optimized using GA, as described in Figure \ref{fig6}.

As discussed in section \ref{GA-PD}, each chromosome represents a problem solution. Every triangular membership has three variable-parameters that can affect the shape; so that each chromosome will have some genes.  Each gene refers to one variable-parameter that affects the triangular shape. Every membership function contains three key ingredients referred as Support, Boundary and Core/Prototype. They can be seen in the Figure \ref{fig121}, whereas their brief definitions summarized are given below:

\begin{figure}[h]
\begin{center}
\includegraphics[width=6cm,height=3.5cm]{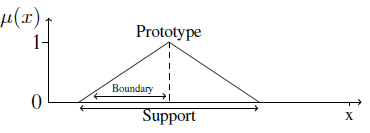} 
\end{center}
\caption{Ingredients of Membership functions.}
\label{fig121}
\end{figure}

\begin{figure}[h]
\centering
\begin{subfigure}{1.0\textwidth}
  \centering
  \includegraphics[width=12cm,height=5.5cm]{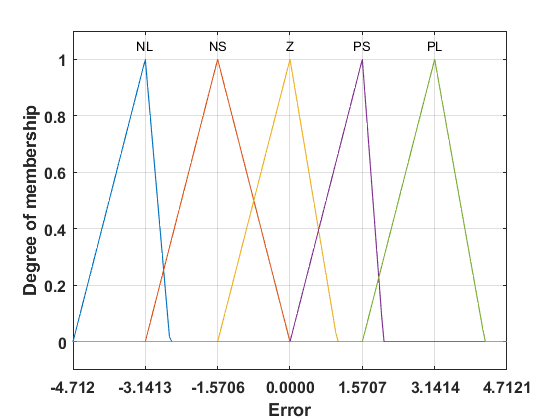}
  \caption{Optimized memberships of the error.}
  \label{fig7}
\end{subfigure}%
\vspace*{8pt}
\begin{subfigure}{1.0\textwidth}
  \centering
  \includegraphics[width=12cm,height=5.5cm]{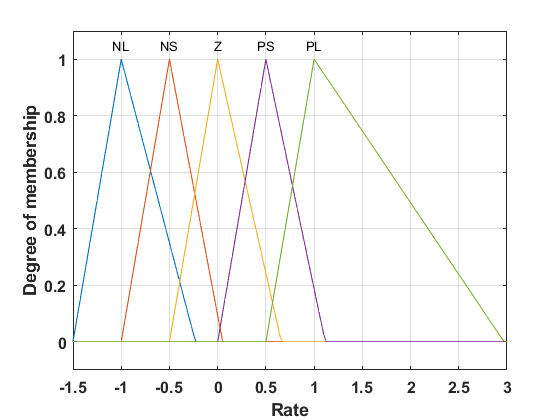}
  \caption{Optimized memberships of the error rate of change.}
  \label{fig8}
\end{subfigure}
\begin{subfigure}{1.0\textwidth}
  \centering
  \includegraphics[width=12cm,height=5.5cm]{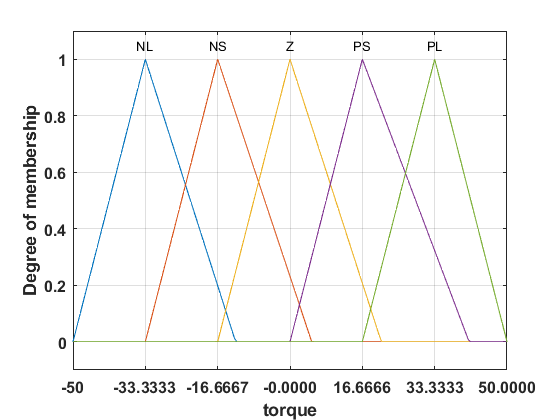}
  \caption{Optimized Memberships of the torque.}
  \label{fig9}
\end{subfigure}%
\caption{Optimized memberships functions of the fuzzy inputs and output. (a): Optimized memberships of the error. (b): Optimized memberships of the error rate of change. (c): Optimized memberships of the torque. }\label{fig30}
\end{figure}

\begin{itemize}
\item Support - The X refers the universe of a given fuzzy variable. Thus for a given fuzzy set A, support refers to that region of the universe X, where all elements x $\varepsilon$ X are characterized by$\mu_{A}$(x) $>$ 0.
\item Core/Prototype - For a given fuzzy set A, core refers to that region of the universe X, where all elements x $\varepsilon$ X are characterized by complete membership of set A, i.e., $\mu_{A}$(x) = 1. The prototype is characterized by the same definition as core but with an exception that there is only one such element where $\mu_{A}$(x) = 1.
\item Boundary - Analogously, the boundary refers to that region of the universe X, where all elements x $\varepsilon$ X are characterized by 0 $<$ $\mu_{A}$(x) $<$ 1.
\end{itemize}
In light of the above definitions and the underlying tuning problem, we are interested in finding out the near optimal values of the above ingredients for each selected variables are at the end of each triangle to maintain the prototype and boundary within its range and to  reduce the computational and complexity of the proposed algorithm. The optimized membership functions for the three linguistic variables at a desired trajectory $\theta_d=[60 \ 50 ]^T$ are as shown in Figures \ref{fig7}, \ref{fig8}, and \ref{fig9}.

The values of the selected variables before and after the optimization process of the membership function for the three linguistic variables are presented in  Table \ref{tab2}.

\begin{wstable}[h]
\caption{the selected variables values before and after Optimization process of the membership function for the three linguistic variables.}
\begin{tabular}{@{}cccccc@{}} \toprule
Linguistic  variables  & NL right  &NS right  & Z right & PS right & PL right \\
 /Membership function  & Befor/After & Before/After& Before/After &Before/After  &Before/After \\
         
\colrule

Error \hphantom{00} & \hphantom{0} -1.57  /-2.6082   & \hphantom{0} -0.000   /-0.0003 & \hphantom{0} 1.57 /1.0282  &\hphantom{0}3.142 /	2.033 &  \hphantom{0}4.712  / 4.2358 \\

Rate of Error\hphantom{00} & \hphantom{0}-0.5/-0.2303 & \hphantom{0}0.000/   0.0513 & \hphantom{0} 0.5 /    0.6599   &  \hphantom{0}1.0  /1.1152 &\hphantom{0}1.5 /2.9688\\

Torque \hphantom{00} &\hphantom{0}-16.667  /	-12.5373  &  \hphantom{0} 0.000 /   5.0038  & \hphantom{0}  16.667 /21.0309  & \hphantom{0} 33.33 /41.2643  & \hphantom{0}50/50 \\

\botrule
\end{tabular}
\label{tab2}
\end{wstable}

\section{Experimental Results and Discussions}\label{sec.5}  

\begin{figure}[h]
\begin{center}
\includegraphics[width=18cm,height=8cm]{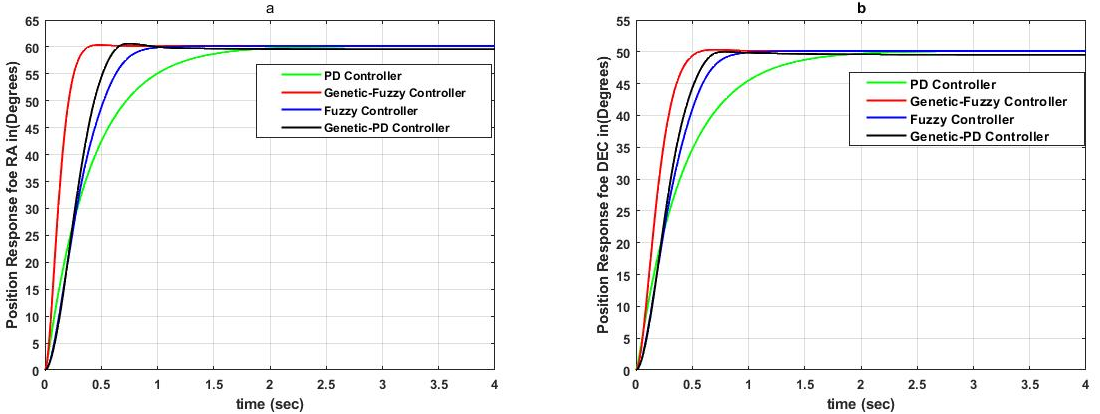} 
\end{center}
\caption{Position responses: (a) RA axis and (b) DEC axis.}
\label{fig12}

\end{figure}

\begin{figure}[h]
\begin{center}
\includegraphics[width=18cm,height=8cm]{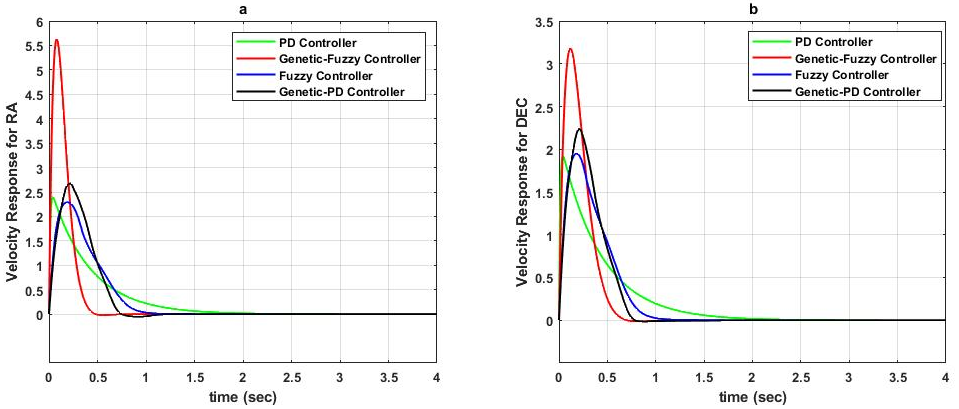} 
\end{center}
\caption{Velocity responses: (a) RA axis and (b) DEC axis.}
\label{fig13}
\end{figure}

The system under study is solved using the Runge-Kutta fifth-order method using the MATLAB Simulink package. This paper studies the change in response of the telescope model with the four mentioned controllers. 
The variations of position and velocity responses for RA and DEC coordinates at the desired trajectory $\theta_d=[60 \ 50 ]^T$ are shown in Figures \ref{fig12} and \ref{fig13}. For all applied controllers, the velocity reaches zero when the telescope reaches the desired trajectory.


The figures show that the proposed GA-FLC provides a better dynamic response compared to the other controllers. The dynamic response of the GA-FLC, GA-PD, FLC, and PD are presented in Table \ref{tab3}.

\begin{wstable}[h]
\caption{Comparisons for system response of proposed controllers in RA and DEC.}
\begin{tabular}{@{}ccccc@{}} \toprule
$Transient response $ & GA-FLC  & GA-PD & FLC & PD \\

                  & RA -  DEC      & RA -  DEC  & RA -  DEC         & RA   - DEC \\
\colrule
Rise Time\hphantom{00} & \hphantom{0} 0.2070 -	0.3045 & \hphantom{0}0.3918	-   0.4129 & \hphantom{0}0.5155 -    0.5040   & 0.8801     -	0.9135 \\

Settling Time \hphantom{00} & \hphantom{0} 0.3314   -	0.4852   & \hphantom{0} 0.3314   -	0.4852   & \hphantom{0}0.8096   -	0.7999  &1.5767 -	1.6366 \\

Overshoot \hphantom{0} & 0  -	0   &  \hphantom{0} 1.7344     -	0.9395 & 0    -	0 \hphantom{0} &  0 -	0 \hphantom{0} \\
\botrule
\end{tabular}
\label{tab3}
\end{wstable}

The results show that the FLC needs 0.5155 and 0.5040 seconds for the RA and DEC positions in response to changes in the desired values, respectively. When the system is attached to the GA-FLC, the results become 0.2070 and 0.3045 seconds for the rise time, which is much better than the FLC. The rise times for the system using classical PD and GA-PD controller are 0.8801 and 0.3918 seconds for the RA responses, and 0.9135 and 0.4129 seconds for the DEC responses, respectively. 

On the other hand, the FLC consumes 0.8096 and 0.7999 seconds for the RA and DEC positions, respectively, to reach the desired values, while the GA-FLC spends 0.3314 and 0.4852 seconds.The settling times consumed by the system using PD and GA-PD controller are 1.5767 and 0.5859 seconds for the RA responses and 1.6366 and 0.6201 seconds for the DEC responses, respectively. 

For the telescope balance, it is impermissible for the response to have an overshoot. Thus, the GA-PD controller is not preferable in the telescope system. The GA-FLC has the least rising time value, settling time compared with the other controllers, besides it has no overshoot. Consequently, the GA-FLC has the best overall performance.

\section{Conclusion}\label{sec.6}

The astronomical observation quality varies with the telescope tracking. This paper discusses the effect of optimizing a fuzzy controller, as well as a PD controller on the 14-inches Celestron telescope tracking. The Genetic algorithm is used to optimize the membership functions of the fuzzy controller, as well as the gains of the PD controller. A performance evaluation of four different controllers is included in this work. The controllers are GA-FLC, GA-PD, FLC, and PD controller.  The results show the effectiveness of the proposed GA-FLC controller to decrease the rise and settling time of the overall system.


\begin{thebibliography}{9}

\bibitem[Attia(1997)]{A.Attia97} Attia, A., [1997] {\it Fuzzy logic control for electric drive of astronomical telescope}, M.Sc. Thesis, Ain Shams University, Faculty of Engineering, Dept. of Electrical power and Machines, Cairo, Egypt ).

\bibitem[Attia(2004)]{A.Attia04} Attia, A., [2004] {\it Experimental Astronomy}, {\bf 18}, 93.

\bibitem[Bachira \& Zoubir(2012)]{Ouamri Bachira12} Bachira, Ouamri \& Zoubir, Ahmed-foitih [2012], {\it  International Journal of Electrical and Computer Engineering (IJECE)} {\bf 2(1)}, 90.

\bibitem[Celestron Operating Manual(1992)]{Celestron92}Celestron [2001]  {\it Operating Manual}, (Celestron International Co.), pp. 1-26 .

\bibitem[Deif {\it et al.}(2011)]{S. Deif11} Deif, S., {\it et al.} [2011] {\it 14th International Conference on Aerospace Sciences \& Aviation Technology} {\bf 14}, 1.

\bibitem[Goldberg(1989)]{D. Goldberg89} Goldberg, D., [1989], {\it Book of Genetic algorithms in search, optimization, and machine learning}, (Addison-Wesley Pub. Co.,United States).

\bibitem[Lewis {\it et al.}(1993)]{F. Lewis93} Lewis, F., {\it et al.} [1993] {\it Book of Control of robot manipulators},  (Macmillan Publishing Co., New York).

\bibitem[Shimojima(1995)]{Koji Shimojima95} Shimojima, K., {\it et al.} [1995], {\it Fuzzy Sets and Systems} {\bf 71(3)}, 295.

\bibitem[Santibaiiez(1998)]{V. Santibaiiez98} Santibaiiez, V., {\it et al.} [1998], {\it Proceedings of the 1998 IEEE 
International Conference on Robotics $\&$ Automation}.

\bibitem[Zadeh(1965)]{L. Zadeh65} Zadeh, L., [1965] {\it  Information and Control 8},338.

\end{thebibliography}
\end{document}